\documentclass[twocolumn, article, floatfix, a4paper]{article}

\usepackage{amsmath,amssymb,amsfonts}
\usepackage{graphicx}
\usepackage{epstopdf}
\usepackage{color}
\usepackage{verbatim}
\usepackage{subfigure}
\usepackage{balance}
\title{Spectroscopic investigation of local mechanical impedance of living cells} 

\author{\normalsize{Luca Costa$^{1,2}$, Mario S. Rodrigues$^3$, N{\'u}ria Benseny-Cases$^4$, V{\'e}ronique Mayeux$^1$, Jo\"{e}l Chevrier$^{5,6}$, Fabio Comin$^1$} \\
\begin{minipage}{.80 \linewidth}
\small
$^{1)}$ \textit{European Synchrotron Radiation Facility, 6 rue Jules Horowitz BP 220, 38043 Grenoble Cedex, France} \\
$^{2)}$ \textit{Universit\'{e} Joseph Fourier BP 53, 38041 Grenoble Cedex 9, France} \\
$^{3)}$ \textit{CFMC/Dep. Fisica, Faculdade de Ci\^encia, Universidade de Lisboa, Campo Grande, 1749-016 Lisboa, Portugal} 	\\
$^{4)}$  \textit{Astbury Centre for Structural Molecular Biology, Leeds University, Leeds, UK} \\ 
$^{5)}$ \textit{CNRS, Inst NEEL, F-38042 Grenoble, France}
\\
$^{6)}$ \textit{Universit\'{e} Grenoble Alpes, Inst NEEL, F-38042 Grenoble, France}
\\ 
\end{minipage} \\
\begin{minipage}{.85 \linewidth}
\small
The mechanical properties of PC12 living cells have been studied at the nanoscale with a Force Feedback Microscope using two experimental approaches. Firstly, the local mechanical impedance of the cell membrane has been mapped simultaneously to the cell morphology at constant force.
As the force of the interaction is gradually increased, we observed the appearance of the sub-membrane cytoskeleton.
We shall compare the results obtained with this method with the measurement of other existing techniques.
Secondly, a spectroscopic investigation has been performed varying the indentation of the tip in the cell membrane and consequently the force applied on it.\\
In contrast with conventional dynamic atomic force microscopy techniques, here the small oscillation amplitude of the tip is not necessarily imposed at the cantilever first eigenmode. This allows the user to arbitrarily choose the excitation frequency in developing spectroscopic AFM techniques.
The mechanical response of the PC12 cell membrane is found to be frequency dependent in the 1 kHz - 10 kHz range. 
The damping coefficient is reproducibly observed to decrease when the excitation frequency is increased.
\end{minipage}
}

\date{\today}

\begin{document}
\maketitle
\epstopdfsetup{outdir=./}
\section{\label{sec:level1}Introduction:}
Atomic Force Microscopes (AFMs) are intensively used for studying cells.
Because they provide the morphology of the specimens at the nanoscale, AFMs have been employed for imaging the cell surface and the submembrane cytoskeleton \cite{braey98,kasas95,LeGrimellec1998695}.
In addition, AFMs are nowadays extensively used for molecular recognition experiments and to explore the energy landscape of receptor-ligand interactions in living cells \cite{dufrene08,muller09}.
Another central aspect is the possibility to apply a force on the cell membrane and measure the cell elasticity.
The mechanical response of the cells is a key observable for diseases diagnostics \cite{Plodinec2012} and cell signaling \cite{engler06,brown09}. More generally, the  elasticity is involved in many of the physiological processes performed by the cell.
Due to the cell viscoelastic behavior \cite{navajas01,navajas03}, the observed mechanical properties could change significantly depending on the frequency probed during an experiment.
In this mainframe we focus on the measurement of the mechanical impedance of PC12 living cells employing atomic force microscopy methods.

In conventional static AFMs, the tip is slowly approached close to the cell membrane and a force vs indentation curve is then recorded. 
Depending on the geometry and the nature of the mechanical contact between the tip and cell, different contact models can be employed to extract the intrinsic elasticity of the cells \cite{navajas05,dufrene06}.
The approach is generally statistic since a set of approach curves needs to be recorded to properly quantify the cell Young modulus.
A two-dimensional force map on the cell may be recorded to get the Young modulus of each cell point \cite{hofmann97,haga00}.
This technique is often time-consuming because a large numbers of force curves needs to be acquired.
Recently, advanced AFM operational schemes have been proposed \cite{garcia_12,natraman2011} for measuring nanomechanical properties of soft samples.
These methods, called multifrequency AFM, allow the user to simultaneously acquire the topography and the mechanical properties of the specimens. This is possible either monitoring the higher harmonics excited while the tip is interacting with the sample \cite{natraman2011}, either by exciting and monitoring the 2$^{nd}$ cantilever eigenmode \cite{garciall}. As a consequence, the sample elasticity is probed at frequencies linked to the cantilever eigenmodes. These techniques are much faster than the conventional force mapping.

We have introduced an alternative operational scheme based on fiber optic detection system, called Force Feedback Microscope (FFM) \cite{io12,costa2013imaging}. It allows the user to simultaneously measure the static force, the elastic force gradient and the damping coefficient, fully characterizing the interaction between the AFM probe and the specimen.
The static force is the output of an active feedback loop that controls the average position of the tip in the space.
A sub-nanometric oscillation amplitude is then imposed to the tip for measuring the force gradient and the damping coefficient.
The use of small oscillation amplitudes indicates the Force Feedback Microscope to be essentially a linear AFM.
At the solid/liquid interface soft cantilevers with a stiffness in the order of 0.01 N/m are employed.
The use of soft cantilevers and small oscillation amplitude is essential to reduce the invasiveness of the AFM experiment \cite{garcia13} when soft samples have to be studied.
A central aspect of the Force Feedback Microscope is the capability to use as feedback signal to record the sample morphology either the force, either the force gradient, either the damping coefficient.
One out of these three different quantities can be arbitrarily kept constant, providing a contrast in the other two physical observables. In these conditions, the contrast is intrinsically dependent on the choice of the feedback signal, the magnitude of the setpoint, the nature of the interaction and the local change of the interaction on the specimens \cite{costa2013imaging}.
An additional key point of the Force Feedback Microscope (FFM) is the possibility to arbitrarily choose the excitation frequency imposed to the cantilever.
This capability is here exploited to characterize the mechanical response of PC12 living cells in the 1 kHz - 10 kHz frequency range.
For this purpose, PC12 living cells have been previously characterized in conventional static mode. Subsequently, cells have been imaged at different constant forces in the FFM mode. The topography is acquired simultaneously to the measured sample stiffness and the damping coefficient.
Finally, a set of indentation curves at different excitation frequencies has been recorded on the top of the PC12 living cells.

\section{\label{sec:level2}Materials and methods:}
\subsection*{Cell line}
PC12 were obtained from ATCC.
PC12 is a cell line derived from a pheochromocytoma of the rat adrenal medulla and is extensively used as a neuronal model. In the experimental conditions used the cells were not differentiated. 

\subsection*{Cell culture}
Some existing protocols have been followed (\cite{viero06,andre03}).
The PC12 cells are cultivated inside 250 ml Flasks produced by \textit{BD Falcon}. The cell medium is a solution of  Minimum Essential Medium, with 10$\%$ Fetal Bovine Serum, supplemented with antibiotics 1$\%$ Penicillin-Streptomycin at 100 ng/ml. All the products are produced by \textit{Life technologies}. The medium is changed every three days. Cells are cultivated inside an incubator at 37$^\circ$C, 5 $\% CO_2$.

\subsection*{Glass support for AFM measurements}
The \textit{glass coverslips} (Thermanox, 13 mm diameter) are previously autoclave-ted for sterilization. The coverslips are deposited inside the 4 wells Culture plates (\textit{Thermo Scientific Nunc dishes IVF}), steryle. 500 $\mu$l of Fibronectin 1$\frac{\mu g}{ml}$ (\textit{Life technologies}) are deposited in each culture plate and the sample is then left incubated for 60 minutes at room temperature. The Fibronectin is then removed and each culture plate is washed 4 times with sterile PBS.\\ 500 $\mu$l of MEM with 10$\%$ Fetal Bovine Serum and 1$\%$ Penicillin-Streptomycin are deposited in each culture plate.

\subsection*{Cells transfer on the glass support for the AFM measurements:}
The buffer inside the flask is removed. A solution of 15 ml PBS, 0.25$\%$ Trypsin at 1 mg/ml is transferred  inside the flask. The flask is then left incubated for 30 minutes at 37$^\circ$C. The cells are now suspended in the solution. 500 $\mu$l of fetal bovine serum are added to the solution. The solution is centrifuged for 5 minutes at 12500 rpm inside a 15 ml centrifuge tube. The solution is removed leaving the cells at the bottom of the tube. 2 ml of MEM, 10$\%$ fetal bovine serum, 1$\%$ Penicillin-Streptomycin at 100 ng/ml are inserted in the tube. Cells are mechanically re-suspended using a Pasteur pipette. The cells concentration is then calculated using a \textit{Neubauer cell}.
Cells have been usually deposited in each culture plate on the top of the coverslip glass at a concentration of 1 million/ml. The cells are then left in the incubator at 37$^\circ$C, 5 $\% CO_2$ for one/two nights and then measured with the FFM. The imaging buffer is MEM, 1$\%$ Penicillin-Streptomycin at 100 ng/ml.
\begin{figure}[htbp]
\begin{center}
\includegraphics[width=\linewidth]{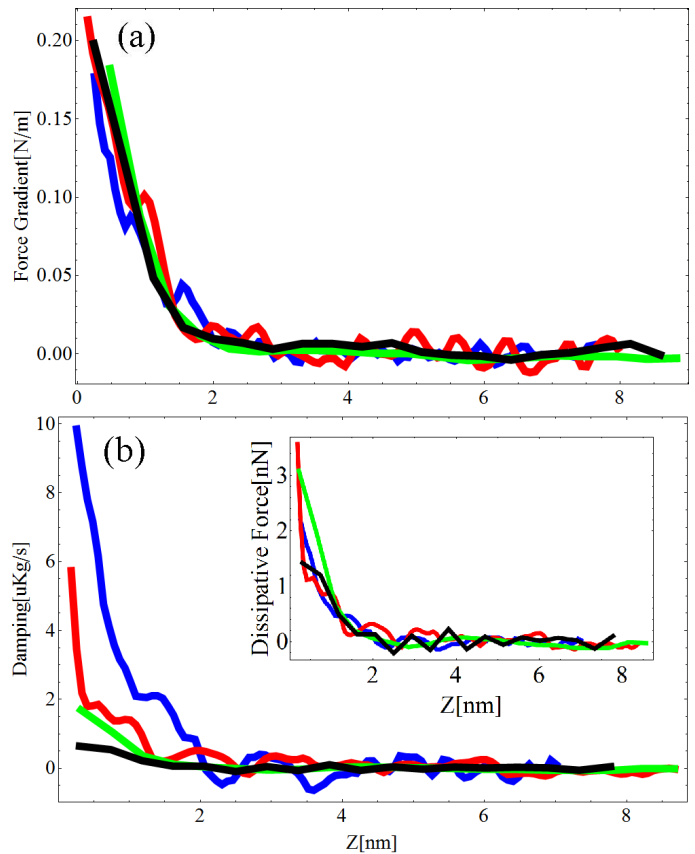}
\caption{
Spectroscopy on the glass in liquid buffer. Force gradient (a) and damping factor $\gamma$ (b) as a function of the excitation frequency. Excitation frequencies: Blue = 1.11 kHz, Red = 3.11 kHz, Green = 9.11 kHz, Black = 11.11 kHz.
}
\label{fig:figure1} 
\end{center}
\end{figure}
\subsection*{Force Feedback Microscopy}
The Force Feedback Microscope is a custom - made AFM based on a fiber optic interferometer detection system \cite{io12}.
Triangular and rectangular MLCT cantilevers provided by \textit{Bruker} with nominal spring constants of 0.01 N/m and 0.02 N/m respectively have been employed. Both the cantilevers have a nominal tip radius of 20 nm.\\
An \textit{Asylum Research} fluid cell has been modified to fit properly in the home made Force feedback Microscope. The glass coverslips have been fixed inside the liquid cell. Cells have been kept at 37$^\circ$C controlling the temperature of the imaging buffer during the measurements.
A central aspect of the Force Feedback Microscope is the calibration of the instrument.
For this purpose, a set of approach curves between the tip and the glass has been recorded for each excitation frequency $\omega$.
\begin{equation}
 \nabla F = a \left[\cos(\phi_\infty)-n \cos(\phi)\right]
 \label{eq:eq1}
\end{equation}
\begin{equation}
 \gamma = \frac{a}{\omega}\left[\sin(\phi_\infty)-n \sin(\phi)\right]
 \label{eq:eq2}
\end{equation}
Equation (1) and (2) convert the measured values $n$ and $\phi$ into the interaction properties $\nabla F$ and $\gamma$.
$\phi$ is the phase between the oscillation of the tip and excitation at the cantilever base, whereas $n$ is the so-called normalized amplitude.
Equation (1) and (2) can be used mainly because the oscillation amplitude imposed to the tip is kept below 1 nm. It follows that the force feedback microscope can be considered as a linear AFM.

In this case, assuming the integrated force gradient to be equal to the measurement of the static force between the AFM probe and the glass, the key point is to determine the parameters $\phi_{\infty}$ and $a$.
The protocol of calibration is explained in details elsewhere \cite{io12}.
In figure \ref{fig:figure1} the elastic force gradient and the damping coefficient measured on the glass at 4 different excitation frequencies are reported.
The elastic force gradient is found to be independent of the frequency probed (figure \ref{fig:figure1}a). The damping coefficient is observed to decrease at larger excitation frequencies (figure \ref{fig:figure1}b).
Since the dissipative force is proportional to the damping coefficient times the speed of the tip, we do not observe a significant decrease in the force needed to maintain the tip oscillations constant when the excitation frequency is increased.
In other words, the dissipative force exerted by the glass on the AFM tip is found to be constant in the 1 kHz - 10 kHz range.
		\begin{figure}
		\begin{center}
  			\includegraphics[width=7cm]{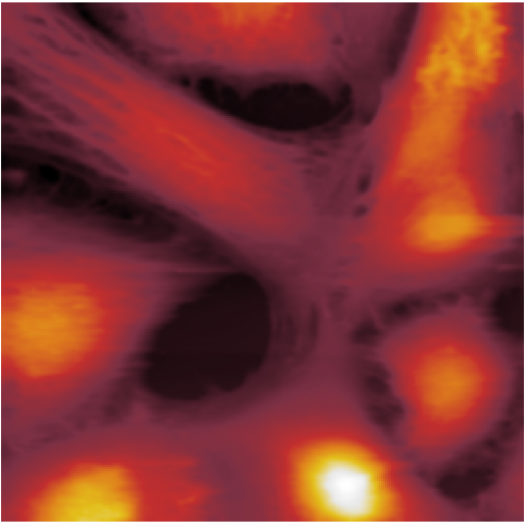}
		\caption{PC12 living cells imaged in conventional contact mode. Scan area = 90 x 90 $\mu m^2$. Several cells of different shapes are here imaged.}
		\label{fig:figure2}
		\end{center}					
		\end{figure}

\section{\label{sec:level3}Results:}
\subsection*{Characterization in conventional static mode}
Cells have been firstly characterized with conventional AFM static mode.
In figure \ref{fig:figure2} a typical image acquired at a constant repulsive force of 100 pN is reported.
Since in our custom - made AFM it is not possible to laterally offset the sample from the tip, a scan area of 100 x 100 $\mu m^2$ is the only sample surface accessible to obtain an image. As a consequence, the cell concentration employed for the cell culture has been refined in order to obtain a cell density on the glass support similar to the one presented in the image.
Indeed, the presence of the glass support in the scan area is a compulsory condition to calibrate the FFM.
	
\begin{figure}
  		\includegraphics[width=\linewidth]{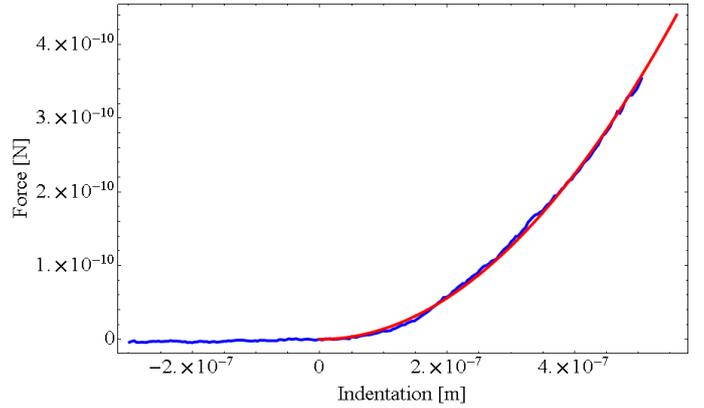}
		\caption{Indentation static curve on a PC12 (blue) fitted with Hertz model(red). The tip is modeled as a four-sided pyramidal indenter.
		The Young's modulus extracted is 1716 Pa using a Poisson ratio equal to 0.5 and $\theta = 25 ^\circ$.}
		\label{fig:figure3}				
\end{figure}

Once the cells are successfully imaged, a set of indentation force curves has been acquired on the living cells.
The tip can be modeled as a four-sided pyramidal indenter. In this case, a relationship between the force $F$ and the indentation $\delta$  \cite{bilodeau92} can be given by 
\begin{equation}
 F = \frac{3 E tan \theta}{4 (1 - {\nu}^2)} {\delta}^2
 \label{eq:eq3}
\end{equation}
where $E$ is the Young modulus and $\nu$ is the Poisson ratio of the cell.
$\nu$ is assumed to be equal to 0.5. The nominal $\theta$ of the probes supplied with the MLCT cantilevers is $ 25 ^\circ$.
In figure \ref{fig:figure3} a typical force vs indentation curve is shown.
The measured Young modulus is 1716 Pa which is comparable to the one measured on PC12 living cells in previous works \cite{chang12}.
\subsection*{Characterization at constant force in force feedback mode}
\begin{figure}[!]
\begin{center}
  		\includegraphics[width=\linewidth]{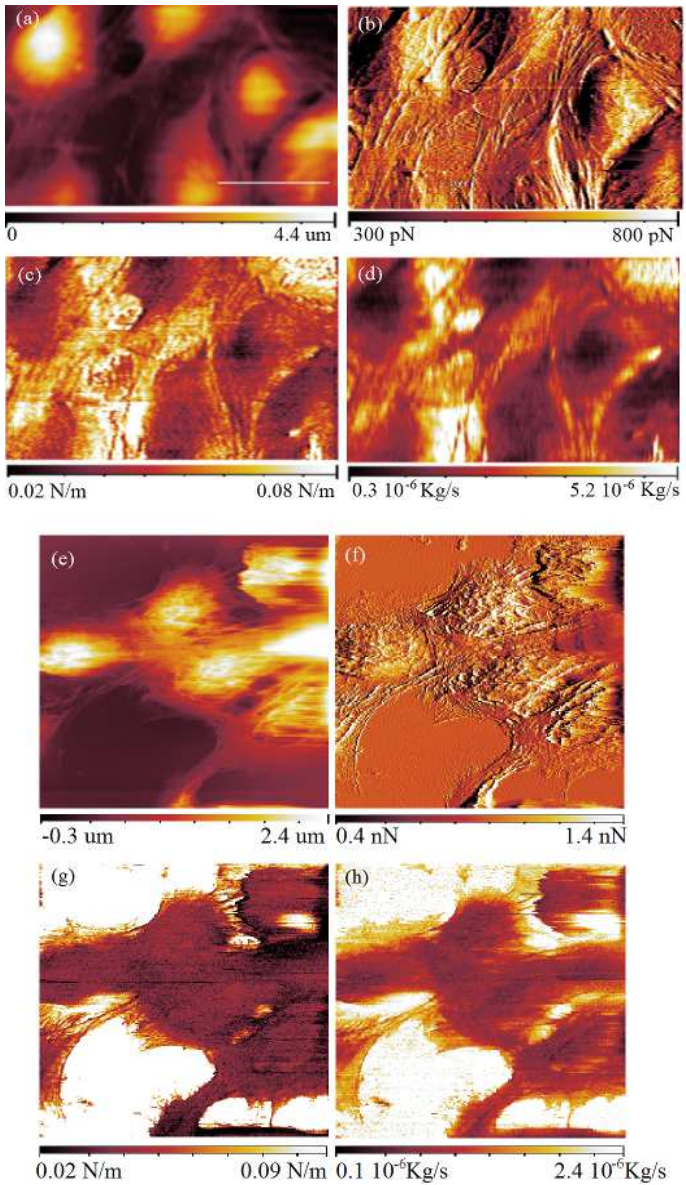}
		\caption{FFM images of PC12 living cells acquired at a constant force of 500 pN and 1 nN. In the images at 500 pN constant force (a,b,c,d) the scale bar is 30 $\mu m$. a) Sample topography, b) force (error), c) force gradient and d) damping factor.\newline
In the images at 1 nN (e,f,g,h) the scale bar is 90 $\mu m$. e) Sample topography, f) force (error), g) force gradient and h) damping factor.}
		\label{fig:figure4}			
\end{center}			
\end{figure}

Cells have been imaged in Force Feedback Microscopy mode at a constant repulsive force of 500 pN.
In this imaging mode, the force is kept constant during the scan and the topography is acquired simultaneously to the force gradient and the damping factor.
An amplitude of 0.4 nm has been imposed to the tip at 2.2 kHz. The Force Feedback Microscope has been then calibrated \cite{io12} with a set of approach curves performed on the glass.
Subsequently, the tip has been brought in contact with the sample at a constant repulsive force of 500 pN and the image has been acquired.
In figure \ref{fig:figure4}a is reported the topography and in figure \ref{fig:figure4}b the force between the tip and the sample which is the error signal. 
In addition, \ref{fig:figure4}c is the measurement of the force gradient, whereas  \ref{fig:figure4}d is the damping factor.\\
The membrane is found to be softer in the center of the cell than at the periphery. 
In analogy with the stiffness, the damping factor is observed to be lower in the center of the cell than at the borders of the cell.
In fibroblasts, these effects can be due to the conformation of the cytoskeleton, in particular of the actin filaments and the microtubules inside the cell as suggested by \cite{haga00,radmacher00,li12}.
These effects can be also due to the higher influence of the glass substrate when the measurements of the stiffness and the damping factor are carried out at the periphery of the cells.
The measured stiffness of the cells is comparable to the stiffness measured in multi-harmonic atomic force microscopy on different cells \cite{natraman2011}.

In a different experiment, PC12 cells have been imaged at the constant force of 1 nN. An amplitude of 0.4 nm has been imposed to the tip at 7.78 kHz. In figure \ref{fig:figure4}e is reported the topography, in figure \ref{fig:figure4}f the force (error signal), in figure \ref{fig:figure4}g the elasticity and in figure \ref{fig:figure4}h the damping coefficient. \\
Figure \ref{fig:figure4}e reveals the presence of actin filaments/microtubules which are however hardly measured in the elasticity and damping coefficient images.
In analogy with the measurement at constant force of 500 pN, stiffness and damping factor are observed to be lower in the center of the cells than at the periphery of the cells.

\subsection*{Characterization at constant force and tunable frequency}
\begin{figure}[!]
  		\includegraphics[width = \linewidth]{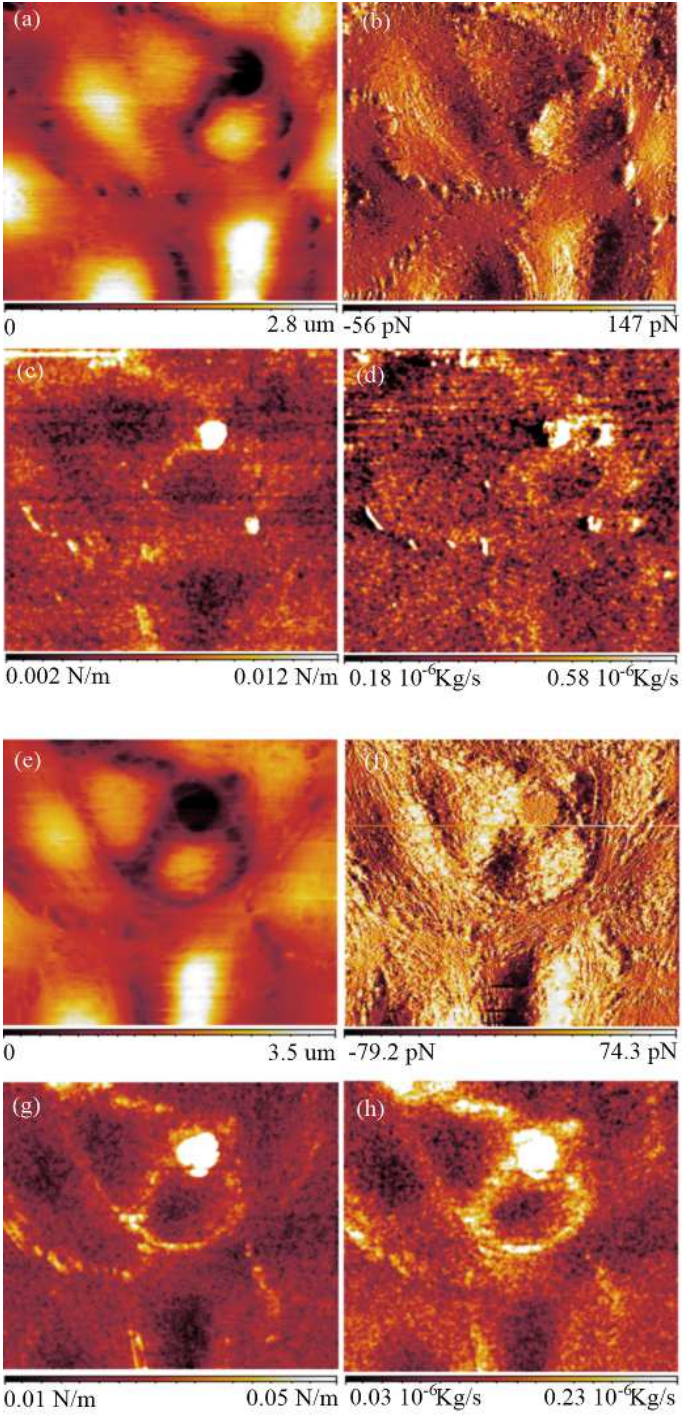}
		\caption{FFM images of PC12 living cells acquired at a constant force of 50 pN. Scan area = 90 x 90 $\mu m^2$. a) Sample topography, b) force (error), c) force gradient and d) damping factor at 2.25 kHz.  e) Sample topography, f) force (error), g) force gradient and h) damping factor at 13.25 kHz}
		\label{fig:figure5}						
\end{figure}

The capability of the FFM to arbitrarily choose the excitation frequency of the AFM tip allows the user to measure the sample force gradient and the sample damping factor at the desirable frequency.
For this purpose a set of calibration curves on the top of the glass substrate have been performed at 2.25 kHz and then at 13.25 kHz. The oscillation amplitude imposed to the tip is 0.4 nm.
Once the FFM is calibrated, one image at constant repulsive force of 50 pN has been acquired at the excitation frequency of 2.25 kHz (figure \ref{fig:figure5}a,b,c,d).
Consequently, the same scan area has been imaged a second time at constant repulsive force of 50 pN, imposing an excitation frequency of 13.25 kHz (figure \ref{fig:figure5}e,f,g,h).
In analogy with the measurement presented in figure \ref{fig:figure4}, the cell center is again observed to be softer than the cell borders.

The stiffness and the damping factor measured at 2.25 kHz are nosier than those measured at 13.25 kHz mainly because of the larger influence of the noise $\frac{1}{f}$. In order to run an acceptable measurement at low frequency, the scan speed at 2.25 kHz has been set twice slower than the one at 13.25 kHz.

The changes in the sample morphology between figure \ref{fig:figure5}a and figure \ref{fig:figure5}e are due to the fact that the cells are alive. We observe consistent differences between the images of the elasticity and the damping factor.
The cells are found to be stiffer at 13.25 kHz than at 2.25 kHz by a factor five.
The damping factor is instead observed to be larger at 2.25 kHz than at 13.25 kHz.
The decrease of the damping factor once the excitation frequency is increased is a behavior observed on the spectroscopy performed on the glass (figure \ref{fig:figure1}).

\subsection*{Indentation curves}
\begin{figure}[htbp]
\centering
  			\includegraphics[width = \linewidth]{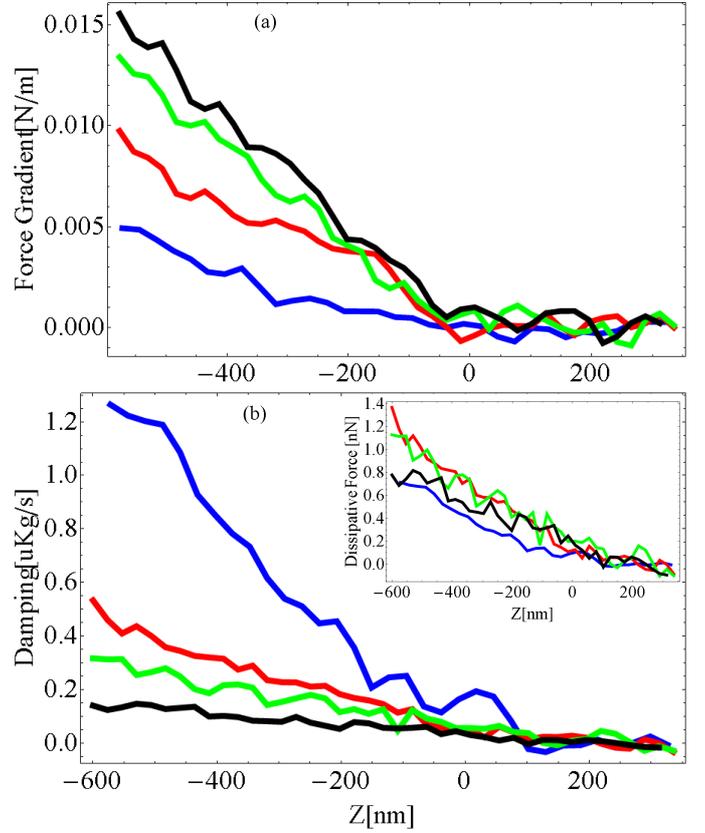}  			
\caption{Spectroscopy of the PC12 in liquid buffer. Force gradient (a), damping factor (b) and dissipative force (inset) as a function of the excitation frequency. Excitation frequencies: Blue = 1.13 kHz, Red = 5.13 kHz, Green = 7.13 kHz, Black = 11.13 kHz.}\label{fig:figure6} 
\end{figure}

A spectroscopic investigation of the PC12 cells has been developed through the acquisition of tip-cell approach curves.
The protocol is the following:
\begin{itemize}
\item The cells were imaged in conventional contact mode. 
\item The Force Feedback Microscope was calibrated with a set of approach curves at different frequencies on the glass. A set of parameters $(\phi_{infty})_i$ and $a_i$ for each excitation frequency $\omega_i$ were obtained. The force gradients $\nabla F_i$ are equal for each excitation frequency as shown in figure \ref{fig:figure1}a.
\item A set of approach curves at different frequency on the PC12 was performed in the highest point of the cell topography.
\end{itemize}
For an applied force of 500 pN, the observed cell elasticity in all the measurements has always been found in the order of 0.01 N/m.
An example of such an experiment is presented in figure \ref{fig:figure6} 
We notice that the damping rate is decreasing when the excitation frequency increases, whereas the force gradient is increasing with the frequency.
The increase of the elasticity with the frequency has not been observed to be reproducible in all the measurements performed.
In particular, we observe a decrease of the cell elasticity at frequencies larger than 13 kHz.
The decrease of the damping factor with the frequency is, on the contrary, reproducible. The resulting dissipative force, presented in the inset of figure \ref{fig:figure6}b, is observed to be poorly dependent on the frequency.
According to \cite{navajas03}, the damping factor is expected to increase with the excitation frequency in the 1 Hz - 100 Hz range. However, in our case the oscillation amplitude imposed to the tip is 2 orders of magnitude lower than the usual amplitudes used for studying living cells.
Since in the present case the oscillation amplitude is much smaller than the cell thickness, we think that the effect of of what is below the membrane may be smaller when compared to the case were the amplitude of oscillation is larger than to the cell thickness.
In addition, a small oscillation amplitude may induce measurements of stiffness and damping factor which are more localized on the cell membrane than in an experiment where larger oscillation amplitudes are imposed to the tip.

\subsection*{Analysis of elastic properties in dynamic mode:}
\begin{figure}[htbp]
\begin{center}
	\includegraphics[width = \linewidth]{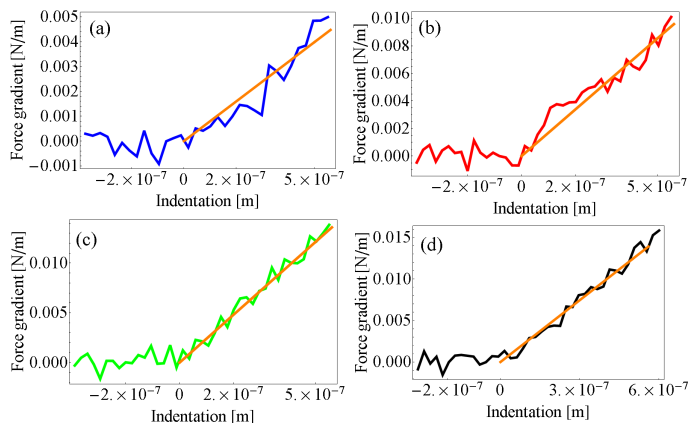}  	
\end{center}		
\caption{Analysis of the PC12 elastic properties through indentations curves. a) Force gradient as a function of the tip indentation. a) $\omega$ = 1.13 kHz, b) $\omega$ = 5.13 kHz, c) $\omega$ = 7.13 kHz, d) $\omega$ = 11.13 kHz. The lines in orange are the experimental linear fit of the cell elasticity using equation \eqref{eq:eq4}.}\label{fig:figure7} 
\end{figure}

Focusing on the elasticity, from equation \eqref{eq:eq3} we can define the force gradient as a function of the tip indentation as
\begin{equation}
\nabla F = \frac{3 E tan \theta}{2 (1 - {\nu}^2)} \delta
 \label{eq:eq4}
\end{equation}
In particular, the data presented in figure \ref{fig:figure6}a have been used to extract the cell Young modulus as a function of the excitation frequency.
The Young modula are found to be equal to 8690 Pa at 1.13 kHz, 18604 Pa at 5.13 kHz, 26385 Pa at 7.13 kHz and 27023 Pa at 11.13 kHz.
The data are reported in figure \ref{fig:figure7}.

In the last decades, the Force Modulation technique \cite{Radmacher1993735} has been applied to the study of the variation of the elasticity of the cells, typically at frequencies lower than 100 Hz \cite{navajas03,Luque20136852}.
In this frequency range the studied cells show an increase of the young modulus with frequency following specific power laws.\\
Here the study has been extended to frequencies larger than than 1 kHz.
As reported in the previous section, no clear reproducibility of the increase of the cell stiffness has been measured.
However, the observed cell elasticity in all the measurements performed has always been found in the order of ten(s) of kPa in this frequency range.
The measured Young modula are therefore observed to be between one and two orders of magnitude larger than the static values (figure \ref{fig:figure3}).

\subsection*{Spatial variation analysis of elastic properties:}
The spatial variation of the cell elasticity can be studied with the FFM during the acquisition of the cell morphology as presented in figure \ref{fig:figure4} and figure \ref{fig:figure5}.
In analogy with the study of the cell elasticity in dynamic mode, it is possible to extract the spatial variation of the Young modulus.
The simultaneous measurement of the elasticity and the applied force allows us to evaluate the Young modulus of the cells through the relation \ref{eq:eq5}.
\begin{equation}
 \frac{(\nabla F)^2 }{F} \approx \frac{3 E tan(\theta)}{1 - \nu^2}
 \label{eq:eq5}
\end{equation}

\begin{figure}[htbp]
\begin{center}
	\includegraphics[width = 7 cm]{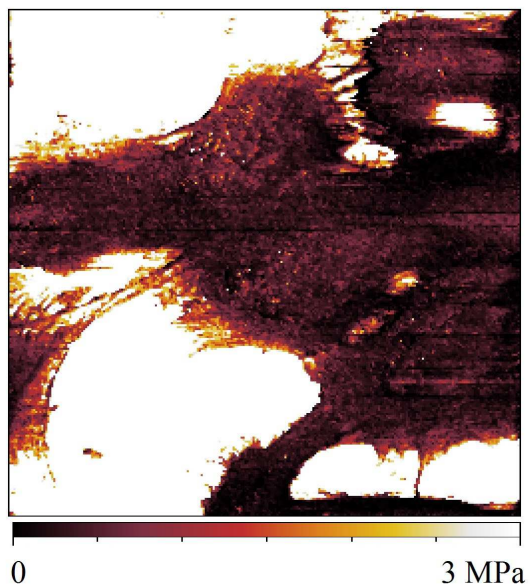}  	
\end{center}		
\caption{Spatial variation of the Young modulus of PC12 living cells imaged at constant repulsive force of 1 nN and at the excitation frequency of 7.78 kHz. The Young modulus is extracted using relation \eqref{eq:eq5}.}\label{fig:figure8} 
\end{figure}

In figure \ref{fig:figure8} the extracted Young modulus of PC12 living cells imaged at a constant repulsive force of 1 nN is reported.

The scale-bar of figure \ref{fig:figure8} is limited to 3 MPa to enhance the contrast in the spatial variation of the cell Young modulus. The periphery of the cells is observed to be much stiffer than the center, reaching values larger than 1 MPa.\\
The Young modulus in the center of the cell is observed to vary from 50 kPa up to 400 kPa. The extracted Young modulus may be compared with the values estimated on different living cells in multi-harmonics AFM \cite{natraman2011}, revealing a consistent agreement.

\section{\label{sec:level4}Conclusion:}
Concluding, the Force Feedback Microscope measures simultaneously the force, the force gradient and the damping factor at different excitation frequencies, fully characterizing the interaction between the AFM tip and the sample. We developed two spectroscopic protocols to study living cells with a Force Feedback Microscope.  The techniques have been applied to the study of PC12 living cells.
At first, cells have been characterized with images at different constant forces and at different excitation frequencies.
Then, the response of the cell membrane has been characterized through the acquisition of approach curves at different frequencies.
The local mechanical impedance, that is the elasticity and the damping factor, have been measured and characterized in the 1 kHz - 10 kHz range.
Our measurements have been compared with data existing in literature revealing that the amplitude of oscillation applied on the tip is a crucial parameter when the cell elasticity and dissipation have to be characterized.
The local mechanical impedance of the PC12 living cell has been characterized in the x/y plane simultaneously to the cell morphology in analogy with already existing technique such as multi-harmonics AFM \cite{natraman2011}.\\
Finally, it is clear that one of the future objectives is the extension of the lower and upper limits of the available excitation frequencies, 1 kHz and 15 kHz respectively.
\vspace{0.5cm}

\noindent \textbf{ACKNOWLEDGMENTS} \\
Luca Costa acknowledges COST Action TD 1002.
Mario S. Rodrigues acknowledges financial support from Funda\c{c}\~{a}o para a Ci\^encia e Tecnologia SFRH/BPD/69201/2010. 
This work was performed at the Surface Science Laboratory of the ESRF and at the PSB laboratories in the Carl-Ivar Br\"{a}nd\'{e}n building in Grenoble.

\balance
\bibliographystyle{ieeetr}
\bibliography{papers}

\end{document}